\documentclass[10pt,fleqn]{article}

\usepackage{amsmath,amsfonts,amssymb,latexsym,cite}
\usepackage{graphicx}

\usepackage{amsfonts,amssymb,cite}
\usepackage{graphicx}

\def\noi{\noindent}

\newcommand{\Title}[1]{\noi {{\Large\bf #1}}\\[1ex]}

\def\Aunames#1{\noi{\bf #1}}
\def\au#1{${}^{#1}$}
\def\Addresses#1{\medskip\noi \protect
	\begin{description}\itemsep -3pt {\it #1} \end{description}}
\def\adr#1#2{\item[${}^{#1}$]{\it #2}}

\def\email#1#2{\footnotetext[#1]{e-mail: #2}\addtocounter{footnote}{1}}

\def\nqq{\hspace*{-2em}}
\def\nhq{\hspace*{-0.5em}}

\def\Jl#1#2{#1 {\bf #2},\ }

\def\ApJ#1 {\Jl{Astroph. J.}{#1}}
\def\CQG#1 {\Jl{Class. Quantum Grav.}{#1}}
\def\DAN#1 {\Jl{Dokl. AN SSSR}{#1}}
\def\GC#1 {\Jl{Grav. Cosmol.}{#1}}
\def\GRG#1 {\Jl{Gen. Rel. Grav.}{#1}}
\def\JETF#1 {\Jl{Zh. Eksp. Teor. Fiz.}{#1}}
\def\JETP#1 {\Jl{Sov. Phys. JETP}{#1}}
\def\JHEP#1 {\Jl{JHEP}{#1}}
\def\JMP#1 {\Jl{J. Math. Phys.}{#1}}
\def\NPB#1 {\Jl{Nucl. Phys. B}{#1}}
\def\NP#1 {\Jl{Nucl. Phys.}{#1}}
\def\PLA#1 {\Jl{Phys. Lett. A}{#1}}
\def\PLB#1 {\Jl{Phys. Lett. B}{#1}}
\def\PRD#1 {\Jl{Phys. Rev. D}{#1}}
\def\PRL#1 {\Jl{Phys. Rev. Lett.}{#1}}

\def\al{&\nhq}
\def\lal{&&\nqq {}}
\def\eq{Eq.\,}

\def\beq{\begin{equation}}
\def\eeq{\end{equation}}
\def\bear{\begin{eqnarray}}
\def\bearr{\begin{eqnarray} \lal}
\def\ear{\end{eqnarray}}
\def\earn{\nonumber \end{eqnarray}}

\def\eqv{\al \equiv \al}

\begin{document}
\twocolumn[

\Title{Accelerating model of flat universe in $f(R,T)$ gravity}

\Aunames{Nishant Singla\au{a,1}, Mukesh Kumar Gupta\au{b,2}, Anil Kumar Yadav\au{c,3}} 

\Addresses{
\adr a {\small Department of Physics, Suresh Gyan Vihar University, Jaipur, India}
\adr b {\small School of Engineering \& Technology, Suresh Gyan Vihar University, Jaipur, India}
\adr c {\small Department of Physics, United College of Engineering and Research, Greater Noida - 201306, India}}


\abstract
{The $f(R,T)$ theory of gravitation is an extended theory of gravitation in which the gravitational action contains both the Ricci scalar $R$ and the trace of energy momentum tensor $T$ and hence the cosmological models based on $f(R,T)$ gravity are eligible to describing late time acceleration of present universe. In this paper, we investigate an accelerating model of flat universe with linearly varying deceleration parameter (LVDP). We apply the linearly time varying law for deceleration parameters that generates a model of transitioning universe from early decelerating phase to current accelerating phase. We carry out the state-finder and Om(z) analysis, and obtain that LVDP model have consistency with astrophysical observations. We also discuss profoundly the violation of energy-momentum conservation law in $f(R,T)$ gravity and dynamical behavior of the model.}\\

\textbf{Kewwords:} Cosmological parameters; Modified theory of gravity; LVDP law; Energy conditions.\\

    
]
\email 1 {nishantsinglag@gmail.com}
\email 2 {mkgupta72@gmail.com}
\email 3 {abanilyadav@yahoo.co.in}

{ 
\def\mn{_{\mu\nu}}
\def\MN{^{\mu\nu}}
\def\mN{_\mu^\nu}
\def\nM{_\nu^\mu}
\def\cK{{\cal K}}
\def\cV{{\cal V}}
\def\eqv{\al \equiv \al}
\def\kappa{\varkappa}

\def\wt{\widetilde}
\def\tg{{\wt g}}
\def\tR{{\wt R}}

\def\M{{\mathbb M}}
\def\N{{\mathbb N}}
\def\R{{\mathbb R}}
\def\S{{\mathbb S}}
\def\V{{\mathbb V}}
\def\oR{{\overline R}}

\def\rf{\eqref}
\def\eqn{\eq\eqref}

\def\bh{black hole}
\def\bhs{black holes}
\def\Swz{Schwarz\-schild}

\vspace{5cm}

\section{Introduction}
\label{intro}
The recent observational data \cite{Peebles/2003} on the late time acceleration of the universe and the existence of dark matter have posed a fundamental theoretical challenge to gravitational theories. However the idea of modification of general relativity was not come to exist just after the discovery of accelerating universe. Several modified theories of gravity such as Brans-Dike theory, scalar-tensor theory etc exist since a long time due to combined motivation coming from cosmology and astrophysics. After discovery of accelerating expansion of universe, the attention of researchers towards modified gravity have been sought and the possibility that the modification of general relativity at cosmological scales can explain dark energy and dark matter becomes an active area of research since 2003~\cite{Nojiri/2003,Capozziello/2003,Capozziello/2008,Nojiri/2011,Singh/2018}. Harko et al.\cite{Harko/2011} have proposed a general non-minimal coupling between matter and geometry by considering the effective gravitational Lagrangian consisting of an arbitrary function of R and T where R and T denote the Ricci scalar and trace of energy-momentum tensor. Thus, in $f(R,T)$ gravity, authors have justified, T as an argument for the Lagrangian from exotic imperfect fluids. Therefore the new matter and time dependent terms in gravitational field behaves like cosmological constant. Thus the extra acceleration arises in $f(R,T)$ gravity is not only due to geometry of space-time but also from the matter content of universe. This extraordinary features of $f(R,T)$ theory of gravitation has attracted many researchers to study and reconstruct this theory in various contexts of astrophysics and cosmology \cite{Sharif/2012,Shabani/2013,Jamil/2012,Houndjo/2012,Singh/2014,Yadav/2019,Zubair/2016,Singh/2016a,Sahoo/2016a,Moraes/2017a}. Some interesting applications of $f(R,T)$ gravity have been given in  References \cite{Yadav/2014,Yadav/2018}. Also, in this conection, Nojiri et al. \cite{Nojiri/2017} have described inflation, bounce and late time evolution in modified theory of gravity. Recently, Yadav et al \cite{Yadav/2019mpla} and Bhardwaj et al \cite{Bhardwaj/2019} have investigated the Bulk viscous embedded cosmological models in $f(R, T) = f1(R) + f2(R)f3(T)$ gravity.\\

Today, it is well known that expansion of current universe is in fact accelerating and it had evolved from decelerating expansion to accelerating expansion. However, we still have no satisfactory explanation for this fact that occur at energy scales $\simeq 10^{-4}$ eV, where we supposedly know physics very well 
\cite{Akarsu/2014a}. It is customary to note that the current accelerated expansion of universe, essentially requires either the presence of an energy source in the context of General Relativity (GR) whose energy density decreases very 
slowly with the expansion of universe \cite{Bamba/2012} or a modification of GR for describing gravitation at cosmological scales \cite{Capozziello/2011a}. Firstly, \cite{Akarsu/2012} have proposed linearly varying deceleration parameter (LVDP)law and later on, \cite{Akarsu/2014} have constrained cosmological parameter of LVDP universe with H(z) + SN Ia data points. This study reveals that the LVDP model is only unbiased model about the future of universe which ends with a big rip. Motivated from studies mentioned above, we assume that the deceleration parameter varies linearly with time to derived the accelerating universe within the frame work of $f(R,T)$ gravity. We followed here the different mechanism with modification in GR whereas      \cite{Akarsu/2012} had produced LVDP cosmological model in GR. It is worth to note that with aid of T i.e. modification term of $f(R,T)$ theory of gravitation, EOS parameter, statefinder parameters are evolving within the range matches with recent observations. Recently, Moraes et al. \cite{Moraes/2018} have given a new approach for the conservation of energy momentum tensor in $f(R,T)$ theory by choosing ordinary matter content. Thus there is need to explore $f(R,T)$ theory with scalar field however the theoretical and observational investigation of scalar field models is a challenging task in cosmology. It is worth to note that $f(R,T)$ theory is also applicable to describe the effects of the modification of Einstein gravity in
the formulation of structure scalars. Some important applications and existence of strange stellar/compact objects within framework of $f(R,T)$ gravity are given in references \cite{Yousaf/2016,Yousaf/2018,Yousaf/2016a,Yousaf/2019}. It is important to note that $f(R,T)$ gravity is gravitationally responsible for mechanism of particle production \cite{Moraes/2018,Harko/2014,Bertolami/2007,Harko/2015,Zaregonbadi/2016,Shabani/2007}. Recently Harko et al \cite{Harko/2015} have showed that is a phenomenon of transforming energy into momentum and vise-versa. \\

In the present work, we extend the work carried out by Akarsu and Dereli \cite{Akarsu/2012} by taking into account, the modification of GR. We investigate the $f(R,T) = f(R) + 2f(T)$ gravity model with linearly varying deceleration parameter in FRW space-time. The paper is organized as follows: in section 2, we present the basic mathematical formalism of $f(R,T) = f(R) + 2f(T)$. In section 3, we have computed the physical and geometrical parameters of derived model. Section 4 deals with the validation/violation of energy condition. In section 5, we have discussed the violation of energy momentum conservation in $f(R,T) = f(R) + 2f(T)$ theory of gravitation. In section 6 we have checked the viability of derived model through the analysis of statefinder parameters, Om(z) parameter, stability of derived solution and jerk parameter. The sum up of findings are accumulated as conclusion in section 7.\\
\section{The $f(R,T) = f(R) + 2f(T)$ formalism}
The geometrically modified action in $f(R,T) = f(R) + 2f(T)$ theory of gravitation is read as
\begin{equation}
\label{eq1}
S = \frac{1}{2\kappa}\int [f(R)+2f(T)]\sqrt{-g}d^{4}x + \int L_{m}\sqrt{-g}d^{4}x
\end{equation}  
where $f(R)$ and $f(T)$ are an arbitrary function of Ricci scalar R, and of the trace T of the stress-energy tensor of the matter $T_{ij}$ and $\kappa = \frac{8\pi G}{c^{4}}$.\\
Here, the energy momentum tensor for perfect fluid distribution, $T_{ij} = -pg_{ij}+(\rho + p)u_{i}u_{j} $ is derived from the matter lagrangian $L_{m}$.  Following, Harko et al. \cite{Harko/2011}, we choose the matter Lagrangian as $L_{m} = -p$, $f(R) = R$ and $f(T) = \lambda T$; $\lambda$ being the constant. p and $\rho$ are the isotropic pressure and energy density respectively. $u^{i}$ is the four velocity of the fluid satisfying $u_{i}u^{i} = 1$ in co-moving co-ordinates.\\
Thus the corresponding field equation is read as
\begin{equation}
\label{eq2}
R_{ij}-\frac{1}{2}R g_{ij} = \kappa T_{ij}+ 2\frac{\partial f}{\partial T}T_{ij}+[f(T)+2p\frac{\partial f}{\partial T}]g_{ij}
\end{equation}
In this paper, we consider the flat FRW metric as
\begin{equation}
\label{eq3}
ds^{2} = dt^{2}-a^{2}(t)[dx^{2}+dy^{2}+dz^{2}]
\end{equation}
where $a(t)$ is the cosmic scale factor.\\
In a coming co-ordinate system, the equation (\ref{eq2}) and (\ref{eq3}), read as
\begin{equation}
\label{eq4}
3H^{2}=(1+3\lambda)\rho-\lambda p
\end{equation}
\begin{equation}
\label{eq5}
2\dot{H}+3H^{2}=\lambda\rho-(1+3\lambda) p
\end{equation}
In the above equations, $H = \frac{\dot{a}}{a}$ is Hubble's parameter and overhead dot denotes time derivatives. We have chosen the unit system such that $\kappa = 1$.\\
The deceleration parameter is defined as
\begin{equation}
\label{dp-H}
q = -\frac{\ddot{a}a}{\dot{a}^{2}} = -1-\frac{\dot{H}}{H^{2}} 
\end{equation}
Equations (\ref{eq5}) and (\ref{dp-H}) lead to
\begin{equation}
\label{q-1}
q = -\frac{\lambda \rho}{2H^{2}}+\frac{(1+3\lambda)p}{2H^{2}}+\frac{1}{2}
\end{equation}
From equation (\ref{q-1}), we observe that acceleration in universe is possible when $\frac{\lambda \rho}{2H^{2}} > \frac{(1+3\lambda)p + H^{2}}{2H^{2}}$. Thus the model under consideration is able to describe late time acceleration of universe without inclusion of cosmological constant or dark energy component.\\
\section{Linearly varying deceleration parameter}
Following, Berman \cite{Berman/1983} and Berman and Gomide \cite{Berman/1988} had proposed the special law for Hubble's parameters that yields the constant value of deceleration parameter. This power law cosmology corresponds to accelerating expansion of universe. Later on, some authors have investigated hybrid expansion law that describes the expansion of universe from early deceleration phase to current acceleration phase \cite{Yadav/2013,Akarsu/2014}. In this paper, we assume a generalized linearly varying deceleration parameter \cite{Akarsu/2012} in which the deceleration parameter is not constant.
\begin{equation}
\label{dp}
q  = -kt+m-1
\end{equation} 
where $k \geq 0$ and $m \geq 0$ are constant.\\
Solving equation (\ref{dp}), we obtain
\begin{equation}
\label{scale}
a=a_{1}e^{\frac{2}{m}arctanh\left(\frac{kt}{m}-1\right)}
\end{equation}
where $a_{1}$ is constant of integration.\\

From equation (\ref{dp}), it is clear that if we set initial time $t_{i} = 0$ then $q = m-1$. For $m > 1$, the universe commence with decelerating expansion in initial phase but with the passage of time, the expansion of universe undergoes from deceleration phase to acceleration phase. The dynamics of deceleration parameter $(q)$ and scalae factor $(a)$ versus time are shown in Fig. 1 \& 2 respectively.\\

\begin{figure*}[thbp]
\begin{center}
\begin{tabular}{rl}
\includegraphics[width=0.50\textwidth]{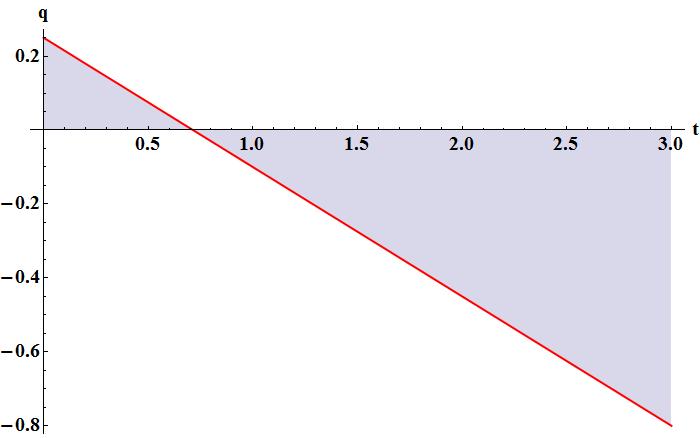}
\end{tabular}
\caption{Dynamics of deceleration parameter.}
\label{fig:1.jpg}
\end{center}
\end{figure*}
The Hubble's parameter is obtained as
\begin{equation}
\label{h}
H = \frac{2}{t(kt-2m)}
\end{equation}
\begin{figure*}[thbp]
\begin{center}
\begin{tabular}{rl}
\includegraphics[width=0.50\textwidth]{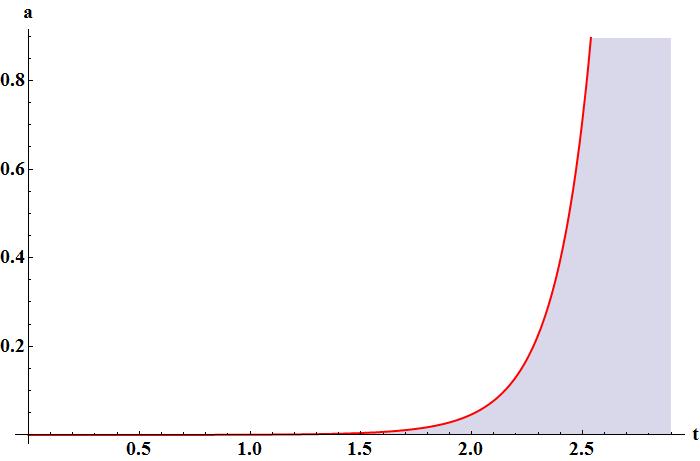}
\end{tabular}
\caption{Plot of scale factor $a$ vs time.}
\label{fig:2.jpg}
\end{center}
\end{figure*}
The energy density and pressure are read as
\[
\rho=\frac{12}{{t^2}{(kt-2m)}^2{(1+4\lambda)}}-\frac{4\lambda}{{(1+2\lambda)}{(1+4\lambda)}}\times
\]
\begin{equation}
\label{rho}
\left[\frac{k}{{t}{(kt-2m)^2}}+\frac{1}{{t^2}{(kt-2m)}}\right]
\end{equation}
\[
p = -\frac{12}{{t^2}{(kt-2m)}^2{(1+4\lambda)}}-\frac{{4}{(1+3\lambda)}}{{(1+2\lambda)}{(1+4\lambda)}}\times
\]
\begin{equation}
\label{p}
~~~~~~~~~~~~~~~\left[\frac{k}{{t}{(kt-2m)^2}}+\frac{1}{{t^2}{(kt-2m)}}\right]
\end{equation}

From Fig. 3, it is clear that initially the energy density is very high and it's value decreases with passage of time. The expression for pressure is given in equation (\ref{p}) and it's behavior is plotted in Figure 4.\\ 

\begin{figure*}[thbp]
\begin{center}
\begin{tabular}{rl}
\includegraphics[width=0.50\textwidth]{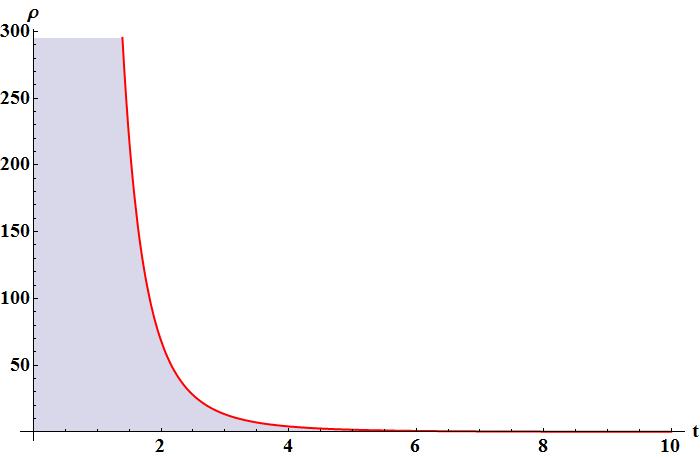}
\end{tabular}
\caption{$\rho$ vs time.}
\label{fig:3.jpg}
\end{center}
\end{figure*}
The density parameter $(\Omega)$ is given by
\[
\Omega=-\frac{1}{(1+4\lambda)}-\frac{{\lambda t^2}{(kt-2m)}^2}{{3}{(1+2\lambda)}{(1+4\lambda)}}\times
\]
\begin{equation}
\left[\frac{k}{{t}{(kt-2m)^2}}+\frac{1}{{t^2}{(kt-2m)}}\right]
\end{equation}
\begin{figure*}[thbp]
\begin{center}
\begin{tabular}{rl}
\includegraphics[width=0.50\textwidth]{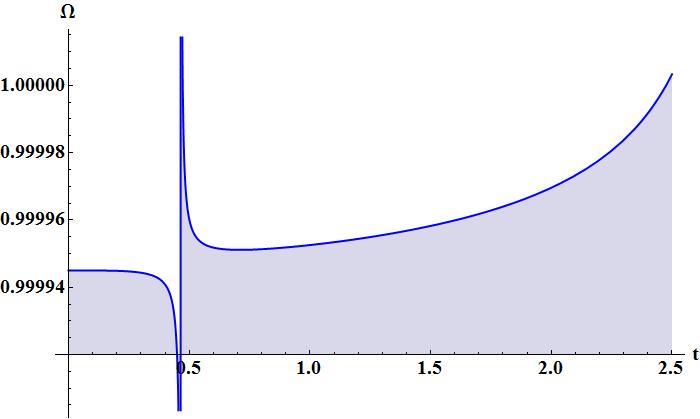}
\end{tabular}
\caption{Density parameter ($\Omega$) vs time.}
\label{fig:5.jpg}
\end{center}
\end{figure*}
The dynamics of density parameter is shown in Fig. 4. It is evident that the overall density parameter behaves like a flip-flop at $t = 0.5$ within narrow interval and approaches 1 for longer times.\\
In order to confront our results with observations, it is customary to obtain the time-redshift relation 
\begin{equation}
\label{tz}
t = \frac{m\left[1+tanh\left(\frac{m}{2}ln\left(\frac{a_{0}}{a_{1}(1+z)}\right)\right)\right]}{k}
\end{equation}
The deceleration parameter $q(z)$ in term of $z$ can be obtained by using the relation $a = \frac{a_{a}}{1+z}$ and (\ref{tz}); $a_{0}$ is the present value of scale factor.
\begin{equation}
\label{q(z)}
q(z) = q_{0}+k~tanh\left(\frac{k+q_{0}+1}{2}ln(\frac{1}{1+z})\right)
\end{equation} 
Here, $q_{0}$ is the present value of deceleration parameter. Fig. 5 depicts the behavior of $q(z)$ with respect to z for $m = 1.25$ and different values of k. The deceleration parameter starts from positive values and evolves up to some negative values $q < -1$ (Fig. 5). Hence the derived model also shows transition from deceleration to acceleration zone for high red-shift values. The value of constant $k$ is calculated by using some observational outcomes \cite{Aviles/2017,Mukherjee/2017,Moresco/2012}. At present (z = 0), the universe is evolving with acceleration. \\
\begin{table*}
\centering
\small
\caption{The computed values of $k$ from observational results}
\vspace{2mm}
\begin{tabular}{@{}crrrrrrrrrrr@{}}
\hline
\hline
$q_{0}$ &~~~~~ Source/Ref.~~~~~ & ~~~red-shift~~~ & k \\
\hline
\hline
-0.5~~~ & Aviles et al. (2017) [JLA + Union 2.1] & z = 0.15~~~ & 0.75 \\
-0.7~~~ & Mukherjee and Banerjee (2017) [OHD + SNe + BAO] & z = 0.29~~~ & 0.95 \\
-0.9~~~ & Moresco et al. (2012) [WMAP + SN Ia] & z = 0.55~~~ & 1.15 \\
\hline
\end{tabular}
\end{table*}
\begin{figure*}[thbp]
\begin{center}
\begin{tabular}{rl}
\includegraphics[width=0.50\textwidth]{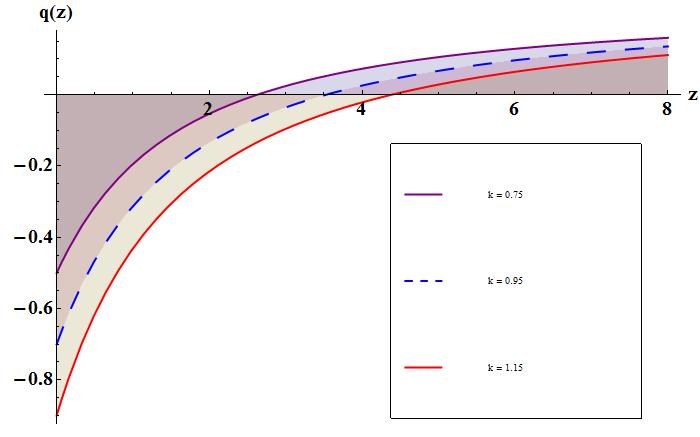}
\end{tabular}
\caption{q(z) versus z for $m = 1.25$ and different values of k.}
\label{fig:7.jpg}
\end{center}
\end{figure*}
\section{Energy conditions}
\begin{figure*}[thbp]
\begin{tabular}{rl}
\includegraphics[width=0.46\textwidth]{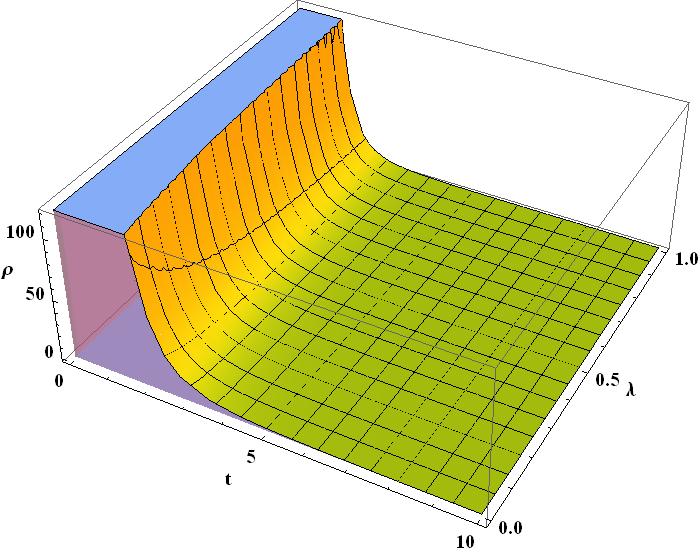}
\includegraphics[width=0.46\textwidth]{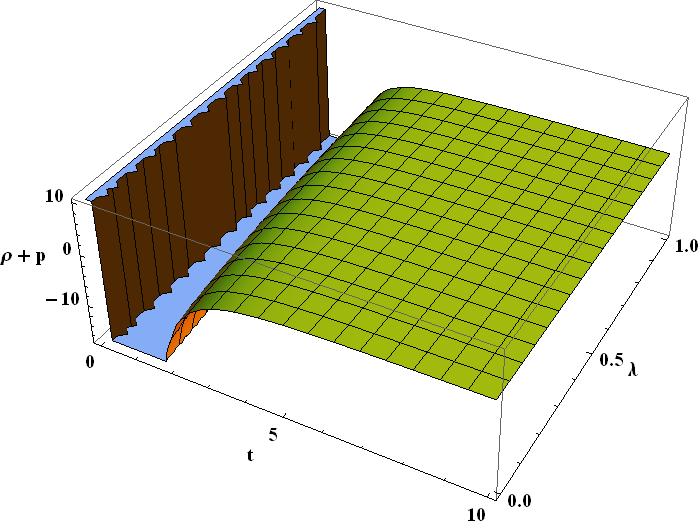}\\
\includegraphics[width=0.46\textwidth]{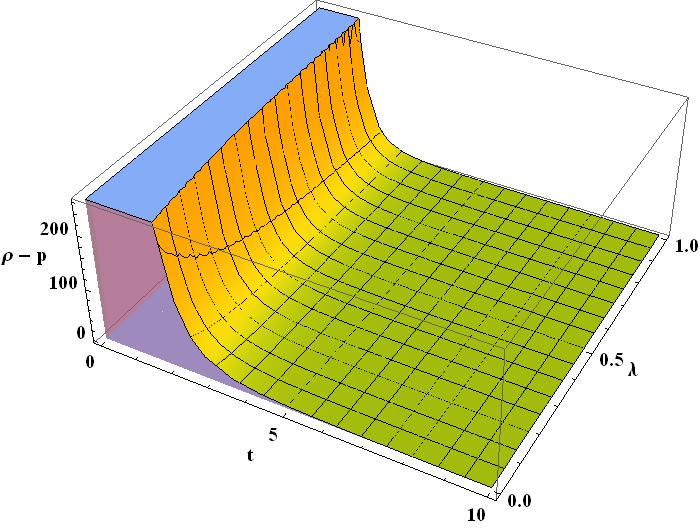}
\includegraphics[width=0.46\textwidth]{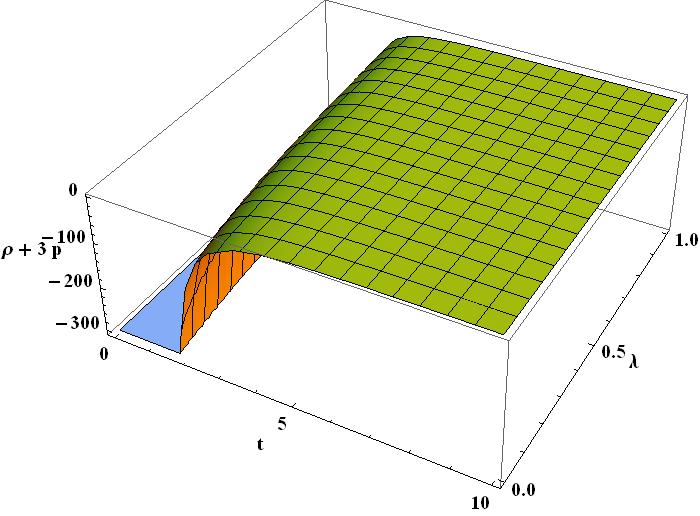}\\
\end{tabular}
\caption{Validation/Violation of energy conditions}
\label{fig:8.jpg}
\end{figure*}
In this section, we will apply the energy conditions to our solution for the energy density and pressure. The main energy conditions in general relativity for the energy-momentum tensor are expressed as\\
Null energy condition $\Leftrightarrow \rho - p \geq 0$\\
Weak energy condition $\Leftrightarrow \rho \geq 0$\\ 
Dominant energy condition $\Leftrightarrow \rho + p \geq 0$\\
Strong energy condition $\Leftrightarrow \rho + 3p \geq 0$\\

Sharif et al.\cite{Sharif/2013} and Alvarenga et al.\cite{Alvarenga/2013} have analyzed the consequences energy conditions in the framework of $f(R,T)$ gravity. The above energy conditions have been graphed in Fig. 6 for $m = 1.25$ and $k = 0.75$. 
\begin{equation}
\rho+p=-\frac{4}{(1+2\lambda)}\left[\frac{k}{{t}{(kt-2m)^2}}+\frac{1}{{t^2}{(kt-2m)}}\right] \leq 0
\end{equation}
\[
\rho-p=\frac{24}{{t^2}{(kt-2m)}^2{(1+4\lambda)}}+\frac{4}{(1+4\lambda)}\times
\]
\begin{equation}
~~~~~~~~~~~~~~~~~~~\left[\frac{k}{{t}{(kt-2m)^2}}+\frac{1}{{t^2}{(kt-2m)}}\right] \geq 0
\end{equation}
\[
\rho+3p=-\frac{24}{{t^2}{(kt-2m)}^2{(1+4\lambda)}}-
\]
\begin{equation}
\frac{{4}{(3+10\lambda)}}{{(1+2\lambda)}{(1+4\lambda)}}\left[\frac{k}{{t}{(kt-2m)^2}}+\frac{1}{{t^2}{(kt-2m)}}\right] \leq 0
\end{equation}
From Fig. 6, we observe that weak energy condition and dominant energy condition are satisfied in the present model while null energy condition as well as strong energy condition are violated. In the derived model, the violation of strong energy condition ensures that anti gravitation effect may be one of the possible cause of acceleration.\\
\section{Violation of energy-momentum conservation law}
For $f(R,T) = R + 2\lambda T$ theory of gravitation \cite{Sahoo/2018}, we have
\begin{equation}
\label{emc-1}
\nabla^{i}T_{ij} = -\frac{2\lambda}{1+2\lambda}\left[\nabla^{i}(pg_{ij})+\frac{1}{2}\nabla^{i}T\right]
\end{equation}
It is important to note that for $\lambda = 0$, $\nabla^{i}T_{ij} = 0$ and one should 
easily retrieves the case of GR. However, in general, for $f(R,T)$ gravity (i.e. $\lambda \neq 0$), the energy- momentum tensor is not conserved. Josset and Perez \cite{Josset/2017} have argued that the non-conservation of energy momentum may arise due to non unitary modifications of quantum mechanics at Plank scale and shown that a non-conservation of energy momentum tensor leads to an effective cosmological constant which decrease with the annihilation of energy during the cosmic expansion and can be reduced to a constant when matter density diminishes. In 2017, Shabani and Ziaie \cite{Shabani/2017} have investigated that the violation of energy - momentum conservation in $f(R,T)$ theory of gravity can provide accelerated expansion. Later on, Shabani and Ziaie \cite{Shabani/2018} have constructed a model of accelerating universe in which an effective field is conserved in $f(R,T)$ gravity rather than the usual energy-momentum tensor. In this paper, we have developed a LVDP model in $f(R,T)$ theory which quantify the violation of energy-momentum conservation through a deviation factor $\Delta$, defined as
\begin{equation}
\label{emc-2} 
\Delta = \dot{\rho}+3H(\rho+p)
\end{equation}
Here, $\Delta \neq 0$.\\
The value of $\Delta$ may be positive or negative depending on weather the energy flows into the matter field either in outward or inward direction respectively. In Fig. 7, we have shown the non-conservation of energy momentum for LVDP law. Fig. 7 depicts that the energy-momentum conservation is validated only for a very limited period of time $(1.1 \leq t \leq 1.3)$. The same behavior is obtained for different values of $\lambda$. 
\begin{figure*}[thbp]
\begin{center}
\begin{tabular}{rl}
\includegraphics[width=0.50\textwidth]{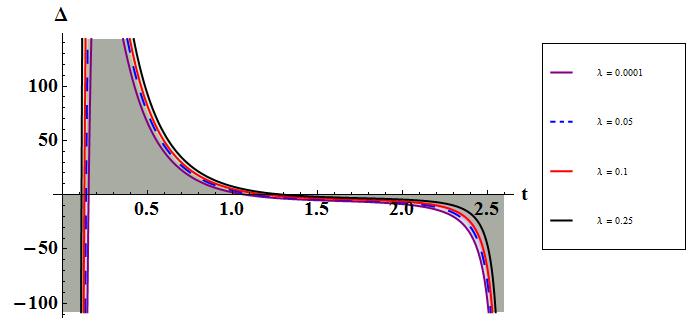}
\end{tabular}
\caption{Non-conservation of energy-momentum.}
\label{fig:7.jpg}
\end{center}
\end{figure*}
\section{Kinematic properties of the model}
\subsection{The statefinder parameters}
The statefinders parameters \cite{Sahni/2003} are obtained as
\[
r=\frac{\dot{\ddot{a}}}{aH^{3}}= \frac{2+4m^{2}+6kt+3k^{2}t^{2}-6m(1+kt)}{2(-2m+kt)^{6}}\times
\]
\begin{equation}
\label{eq1-r}
~~~~~~Exp\left(-\frac{4~Arctanh(1-kt/m)}{m}\right)
\end{equation} 

\[
s=\frac{r-1}{3(q-1)/2}=\frac{-\frac{3}{2}+mt-kt^{2}}{3}\times
\]
\begin{equation}
\label{eq1-s}
\;\;\;\;\;\;\;\;\;\;\;\;\left[-1+g~Exp\left(-\frac{4~Arctanh(1-kt/m)}{m}\right)\right]
\end{equation} 
where $g =\frac{2+4m^{2}+6kt+3k^{2}t^{2}-6m(1+kt)}{2(-2m+kt)^{6}}$.\\

The remarkable feature of statefinders is that these parameters depend on scale factor and its time derivatives and hence are geometric in nature. Fig. 8 exhibits the evolutionary trajectories of derived model in s-r plane. Moreover, the well known flat $\Lambda CDM$ model corresponds to the points $s = 0$ and $r = 1$ in s-r plane. The blue dot in Fig. 8 at (s, r) = (0,1) shows the position of flat $\Lambda CDM$ model.
\begin{figure*}[thbp]
\begin{center}
\begin{tabular}{rl}
\includegraphics[width=0.50\textwidth]{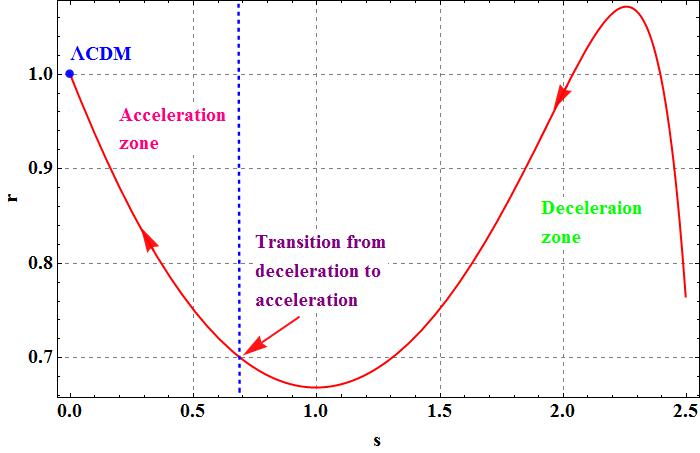}
\end{tabular}
\caption{Dynamics of $r~~\&~~s$.}
\label{fig:7.jpg}
\end{center}
\end{figure*}
\begin{figure*}[thbp]
\begin{tabular}{rl}
\includegraphics[width=0.46\textwidth]{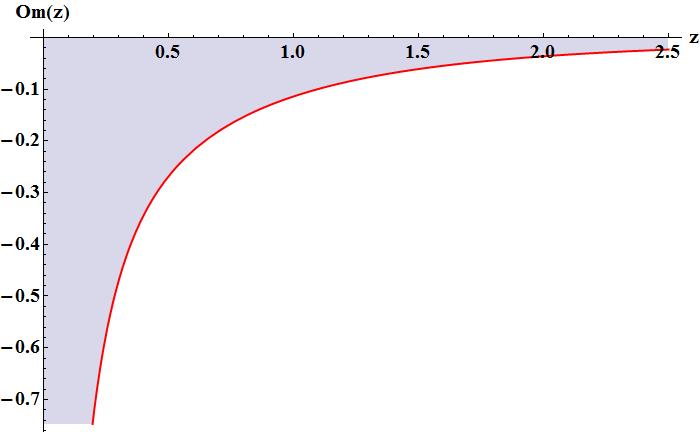}
\includegraphics[width=0.46\textwidth]{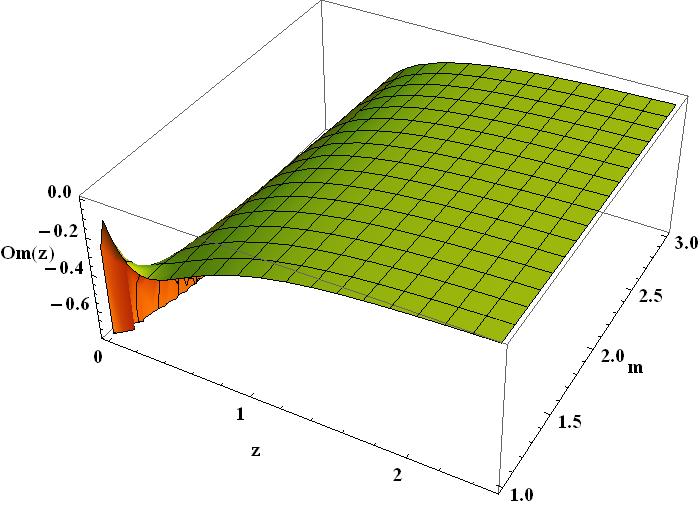}\\
\end{tabular}
\caption{Om(z) parameter versus z for $m = 1.25$ (left panel) and $1 \leq m \leq 3$ (right panel)}
\label{fig:4.jpg}
\end{figure*}
\subsection{Om(z) diagnostic analysis} 
The Om(z) parameter is read as
\begin{equation}
\label{Om-1}
Om(z) = \frac{\left(\frac{H(z)}{H_{0}}\right)^{2}-1}{(1+z)^{3}-1}
\end{equation}
where $H_{0}$ is the present value of Hubble's parameter.
The Om(z) parameter of derived model is given by
\begin{equation}
\label{Om-2}
Om(z)=\frac{\left[\frac{1}{m[1-tanh^{2}[\frac{m}{2}ln(1/1+z)]]}\right]^{2}-1}{(1+z)^{3}-1}
\end{equation}
In the literature, Om diagnostic analysis is useful to modelling the dynamics of dark energy \cite{Sahni/2008}. In comparision with the state finder diagnosis, the Om parameter involves only first derivative of scale factor. The positive, negative and zero values of Om(z) parameter consistent with phantom, quintessence and $\Lambda$CDM dark energy models respectively \cite{Sahooetal/2018}. Fig. 9 depicts the behaviour of Om(z) parameter against z of derived model. The left panel of Fig. 9 shows the dynamics of Om(z) parameter for particular value of $m = 1.25$ whereas the right panel explores the nature of Om(z) parameter in the range $0\leq m \leq 3$. In both the panel, Om(z) parameter is negative and monotonically increasing within the interval $0\leq z \leq 2.5$ which also suggests that at present the universe is in accelerating mode.\\
\subsection{Stability condition}
In this section, we examine the stability of derived model with respect to the perturbation in scale factor as following.
\begin{equation}
\label{stability1}
a\rightarrow a_{B}+\delta a = a_{B}(1+\delta \alpha)
\end{equation}
where $\delta \alpha = \frac{\delta a}{a_{B}}$ denotes the small deviation in perturbed term $\delta a$ and $a_{B}$ is the background scale factor.\\ 
With reference to equation (\ref{stability1}), the perturbations of volume scalar and expansion scalar are read as
\begin{equation}
\label{stability2}
V\rightarrow V_{B}+V_{B}\delta \alpha, \,\,\,\,\,\theta \rightarrow \theta_{B}+\theta_{B}\delta \alpha
\end{equation}
where $V_{B}$ and $\theta_{B}$ denote background volume scalar and expansion scalar respectively.\\

Following, Saha et al (\cite{Saha/2012}), $\delta \alpha$ satisfy the following equations.
\begin{equation}
\label{st6}
\delta\ddot{\alpha}+\frac{\dot{V}_{B}}{V_{B}}\delta\dot{\alpha} = 0
\end{equation}
The background volume scalar is given by
\begin{equation}
\label{st7}
V_{B} =  a^{3} = a_{1}^{3}e^{\frac{6}{m}arctanh\left(\frac{kt}{m}-1\right)}
\end{equation}
Integrating equation (\ref{st6}, we obtain
\begin{equation}
\label{st8}
\delta \alpha = c_{1}- c_{2}e^{\frac{6}{m}arctanh\left(\frac{kt}{m}-1\right)}t
\end{equation}
where $c_{1}$ and $c_{2}$ are the constants of integration.\\

Thus the actual fluctuation in the derived solution is obtained as
\begin{equation}
\label{st9}
\delta a = \delta_{1}e^{\frac{2}{m}arctanh\left(\frac{kt}{m}-1\right)}-\delta_{2}e^{\frac{8}{m}arctanh\left(\frac{kt}{m}-1\right)}t 
\end{equation}
where $\delta_{1} = c_{1}a_{1}^{3}$ and $\delta_{2} =c_{2}a_{1}^{3}$.\\
\begin{figure*}[thbp]
\begin{center}
\begin{tabular}{rl}
\includegraphics[width=0.50\textwidth]{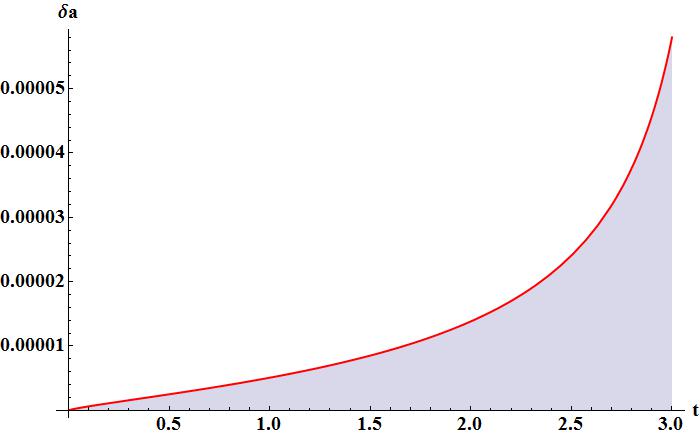}
\end{tabular}
\caption{Plot of $\delta a$ versus t.}
\label{fig:stability.jpg}
\end{center}
\end{figure*}
\begin{figure*}[thbp]
\begin{center}
\begin{tabular}{rl}
\includegraphics[width=0.52\textwidth]{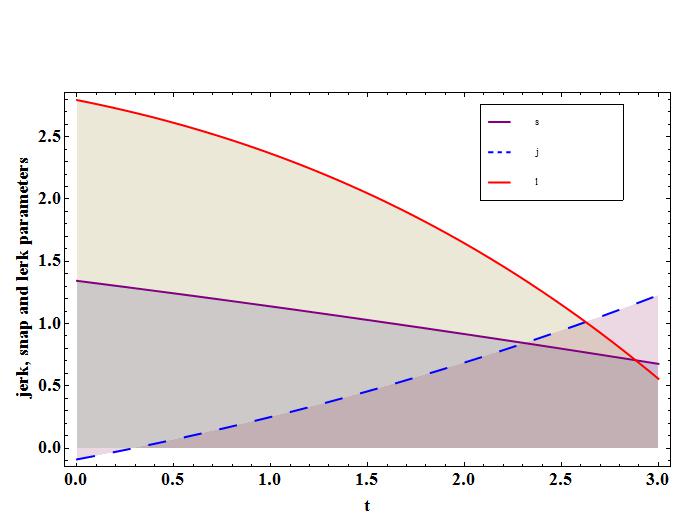}
\end{tabular}
\caption{Plot of jerk, snap and lerk parameter versus t.}
\label{fig:stability.jpg}
\end{center}
\end{figure*}
The straightforward behaviour of actual fluctuation in the derived solution is shown in Fig. 10. We observe that the value of $\delta a$ was null at initial epoch $i. e.$ $t = 0$ and increase slowly with the evolution of universe and finally approaches to a very small positive value. Thus we conclude that the actual fluctuation in derived model is very small which is not desirable to effect the physical properties of universe.\\
\subsection{The jerk, snap and lerk parameters}
The jerk, snap and lerk parameters \cite{Visser/2005,Valent/2018,Visser/2004,Hassan/2011,Hassan/2018} of derived model are obtained as 
\[
j = \frac{\dot{\ddot{a}}}{aH^{3}}
\]
\begin{equation}
\label{jerk}
= -\frac{3 k^2 t^2}{2}+3 m (k t+1)-3 k t-2 m^2-1
\end{equation} 
\[
s = \frac{\ddot{\ddot{a}}}{aH^{4}}= 3 k^3 t^3-3 m \left(3 k^2 t^2+6 k t+2\right)+
\]
\begin{equation}
\label{jerk}
\;\;\;\;\;9 k^2 t^2+m^2 (12 k t+11)+6 k t-6 m^3+1
\end{equation} 
\[
l = \frac{\dot{\ddot{\ddot{a}}}}{aH^{5}}= -\frac{15 k^4 t^4}{2}-30 k^3 t^3-5 m^2 \left(12 k^2 t^2+22 k t+7\right)
\]
\[
\;\;\;\;\; - 30 k^2 t^2+10 m \left(3 k^3 t^3+9 k^2 t^2+6 k t+1\right)+
\]
\begin{equation}
\label{jerk}
\;\;\;\;\;10 m^3 (6 k t+5)-10 k t-24 m^4-1
\end{equation} 
The graphical behaviour of jerk, snap and lerk parameters are shown in Fig. 11. It is important to note that in Ref. \cite{Capozziello/2014}, authors have shown that one can determine the kinematics of universe through the dynamics of $H$, $q$ and $j$. The gold sample of SN Ia observational data predicts the value of jerk parameter as $j = 2.16^{+0.81}_{-0.75}$ \cite{Riess/2004}. Recently, Amirhashchi \& Amirhashchi \cite{Hassan/2018} and Muthukrishna \& Parkinson \cite{Muthukrishna/2016} have estimated the present values of $j$, $s$ and $l$ by using different observational data sets.\\      
\section{Conclusion}
In this paper, we have investigated an accelerating model of flat universe in modified theory of gravity without inclusion of cosmological constant. Therefore, the cosmological constant problems are not associated with the model under consideration. According to the observations and associated analysis, the universe undergoes an accelerated expansion in the present epoch. The universe might have transitioned from early decelerating phase to current accelerating phase which clearly indicate that the evolving deceleration parameter displays a signature flipping behavior that is why, in this paper we have considered LVDP law to construct the model of transitioning universe in $f(R,T)$ gravity. The model straightforward consider the implications of Hubble's parameter, density parameter and deceleration parameter. Collectively, all these confirm the transition of universe and matter-geometry approach clearly indicates the flipping of early deceleration phase to acceleration at present epoch.\\ 

At $t = 0$, both the scale factor and volume becomes zero which means the derived model have point type singularity. 
Initially, $q = m-1$, may     
be positive or negative depending upon the value of m but for non zero cosmic time (i. e. t = m/k), we obtain $q = -1$ which shows asymptotic expansion. The plot of energy density (Fig. 3) illustrates that $\rho$ is very high at initial time and it deceases sharply with passage of time and ultimately approaches to zero at $t\rightarrow \infty $. Fig. 5 exhibits the dynamics of density parameter $(\Omega)$ versus time.\\ 

From Fig 5, it is analyzed that the present value of deceleration parameter is in the range $-0.50 \leq q \leq -0.90$ for $m = 1.25$ and different value of free parameter k. Also we observe that for $k = 0.95$, $q_{0} = -0.70$. This value of $q_{0}$ matches with recent astrophysical observations \cite{Hinshaw/2013}. Hence, for all
graphical analysis we have chosen $k = 0.95$ and $m = 1.25$. The weak and dominant energy conditions are satisfying the model because of the non-increasing and positive energy density which decreases with expansion of universe. The violation of strong energy condition confirms the accelerating expansion of universe. We have discussed the violation of energy momentum tensor for derived model that lead to a sort of accelerated expansion of universe. The $f(R,T)$ theory of gravity explain an accelerating expansion of universe at the cost of non conservation of energy momentum tensor of matter. The state-finder analysis shows that LVDP cosmological 
models approaches the standard $\Lambda CDM$ model in future. We observe that when the redshift z is 
increasing within the interval ~$0 \leq z \leq 2.5$, the Om(z) is monotonically increasing, 
which also indicates an accelerated expansion of universe.\\ 

\section*{Acknowledgments}
The authors are grateful to S. D. Odintsov, Z. Yousaf and B. Saha for fruitful comments on the paper.

\end{document}